\documentclass[%
 reprint,
 amsmath,amssymb,
 aip,
nofootinbib
]{revtex4-2}

\usepackage{graphicx}
\usepackage{dcolumn}
\usepackage{bm}

\usepackage{endnotes}
\begin{document}


\title{Symmetry's Edge in Cortical Dynamics \\ Multiscale Dynamics of Ensemble Excitation and Inhibition}

\author{Nima Dehghani}
 \email{nima.dehghani@mit.edu}
\affiliation{Department of Physics, MIT, Cambridge, MA}
\affiliation{%
 McGovern Institute for Brain Research, MIT, Cambridge, MA
}%
\thanks{Current address: McGovern Institute for Brain Research, MIT, Cambridge, MA\textsuperscript{\dag}}

 \homepage{http://compneuro.mit.edu/}

\date{\today}

\begin{abstract}
Creating a quantitative theory for the cortex presents challenges and raises questions. What are the significant scales--micro, meso, or macroscopic? What are the interactions--pairwise, higher order, or mean-field? And what control parameters are relevant--noisy, dissipative, or emergent?

We suggest an approach inspired by advances in understanding matter. This involves identifying invariances in neuron ensemble dynamics, searching for order parameters connecting key degrees of freedom and distinguishing macroscopic states, and pinpointing broken symmetries to uncover emergent laws when neurons interact and coordinate.

Using multielectrode and multiscale neural recordings, we measure population-level ensemble Excitation/Inhibition (E/I) balance, differing from the input-level E/I balance of single neurons, to study collective behavior in large neural populations. We investigate a set of parameters that can assist us in differentiating between various functional system states (during wake/sleep cycle) and pinpointing broken symmetries that serve different information processing and memory functions. Furthermore, we identify pathological broken symmetries that result in states like seizures.

\end{abstract}

\maketitle


\section*{Introduction}
Over the past two decades, advances in microelectronic fabrication techniques have made it possible to record the electrical activity of large populations of neurons both \emph{in vitro} and \emph{in vivo}. Early studies of neural coding focused on the responses of individual neurons to stimuli \citep{Hubel_1959,Mountcastle_1957}, but modern technology has enabled a shift attention toward population dynamics and pairwise and/or higher-order correlations among neuron ensembles \citep{Churchland_2012,Mattia2002,Averbeck2006,Shimazaki2012,Staude2010,Cohen2011,de2007,Kirkby2013}. While it is widely agreed that information is processed and represented through the correlated activity of large neuron populations, the existing descriptive statistical models often fall short in providing deep theoretical insights into the collective behavior of neural populations. 

Attempts to explain the high variability and spontaneous activity seen in such populations pose significant modeling challenges, pushing the investigators to take a top-down compartmentalization approach \citep{Maass_2016}. This top-down view, loosely follows Marr’s tri-level hypothesis, which separates neurobiological systems into computational, algorithmic, and physical implementation levels \citep{Marr1976,Marr_vision2010}. However, the utility of this separation in neuroscience has recently been debated, for and against \citep{Lengyel2024,Pillow2024}. For refinement, some have suggested to start from the tri-level approach and add further requirements such as hierarchical representation and learning \cite{Valiant_2014}. The core issue in this case is the assumption of separation of scales which contradicts the interconnected nature of cortical circuitry and its functional dynamics. 

Alternatively, there is a need for a more integrated, quantitative, and computational framework that acknowledges the presence of multi-scale and potentially non-separable interactions \citep{Dehghani_2018}. For instance, cortical dynamics likely involve interwoven spatial and temporal scales, where local sub-assemblies link micro- and meso-level phenomena to macro-level behaviors \citep{Dehghani_2018,Simon1962}. Traditional methods like correlation measures or dimensionality reduction of population dynamics often prove inadequate in this context. 

A key example of such multi-scale interactions involves the notion of excitation-inhibition (E/I) balance. One line of research focuses on \emph{``input''} E/I balance—the matching of excitatory and inhibitory inputs at the single-neuron level \citep{Renart2010,vanVreeswijk1996}. This ensures stable firing rates, preventing runaway activity. A complementary perspective, however, centers on \emph{``ensemble''} E/I balance, which aggregates the spiking of large populations of excitatory and inhibitory neurons. Notably, our previous work \citep{Dehghani2016} has highlighted that aggregated excitatory and inhibitory firing rates remain balanced across many conditions, including awake and sleep states. We have also shown that classical maximum-entropy models (e.g., Ising-type models) capture inhibitory spiking patterns well but fail to fully explain excitation \citep{Zanoci2019}, hinting that more nuanced frameworks may be necessary.

Technological limitations introduce additional complications: most current recording devices massively subsample from a cortical column \citep{Meyer_2010,Rakic_2008,Herculano_Houzel_2008,Rakic_1988}, making it challenging to map local population recordings to genuine meso- or macro-scale phenomena. Nonetheless, localized data have already revealed patterns such as stereotypical traveling waves \citep{Le_Van_Quyen_2016} and wave-like spatial propagation \citep{Galinsky_2020}, reinforcing the possibility that multi-scale E/I balance might be part of a broader organizing principle. 

From a broader theoretical lens, statistical physics offers powerful concepts such as \emph{invariances}, \emph{order parameters}, and \emph{broken symmetries}, which can clarify how large-scale organization emerges from microscopic interactions. Insights from physics become particularly valuable here, especially when focusing on identifying patterns and symmetries. Such patterns can uncover order parameters and broken symmetries, offering clues to the underlying laws governing the system \citep{Sethna1992,Sethna2006}. Inspired by renormalization group ideas, we can look for scale invariance in E/I balance by successively coarse-graining the neuronal population activity and observing if certain statistical properties remain unchanged. Such approaches can help identify potential “fixed points” of the system and deepen our understanding of how cortical networks maintain robust E/I balance across diverse spatial or temporal scales \citep{Niedermeyer2005,Nunez2006,Pettersen2008,Bedard2006,Linden2011,Dehghani2012,Zanoci2019}.

Moreover, the search for scale-free features has been connected to the idea of \emph{criticality}, wherein neural systems operate near a phase transition, potentially optimizing computational capabilities \citep{Papanikolaou2011,Sethna2001,Friedman2012}. Rather than focusing solely on cognition or behavior as metrics for describing the system, a more effective approach may involve identifying invariances in ensemble dynamics, seeking order parameters that connect critical degrees of freedom to differentiate macroscopic states, and exploring broken symmetries in the order parameter space. This approach helps uncover the emergent laws governing the coordination and interaction of neurons at scale.

\begin{widetext}
\onecolumngrid
\begin{figure}[ht]
\includegraphics[scale=0.55]
{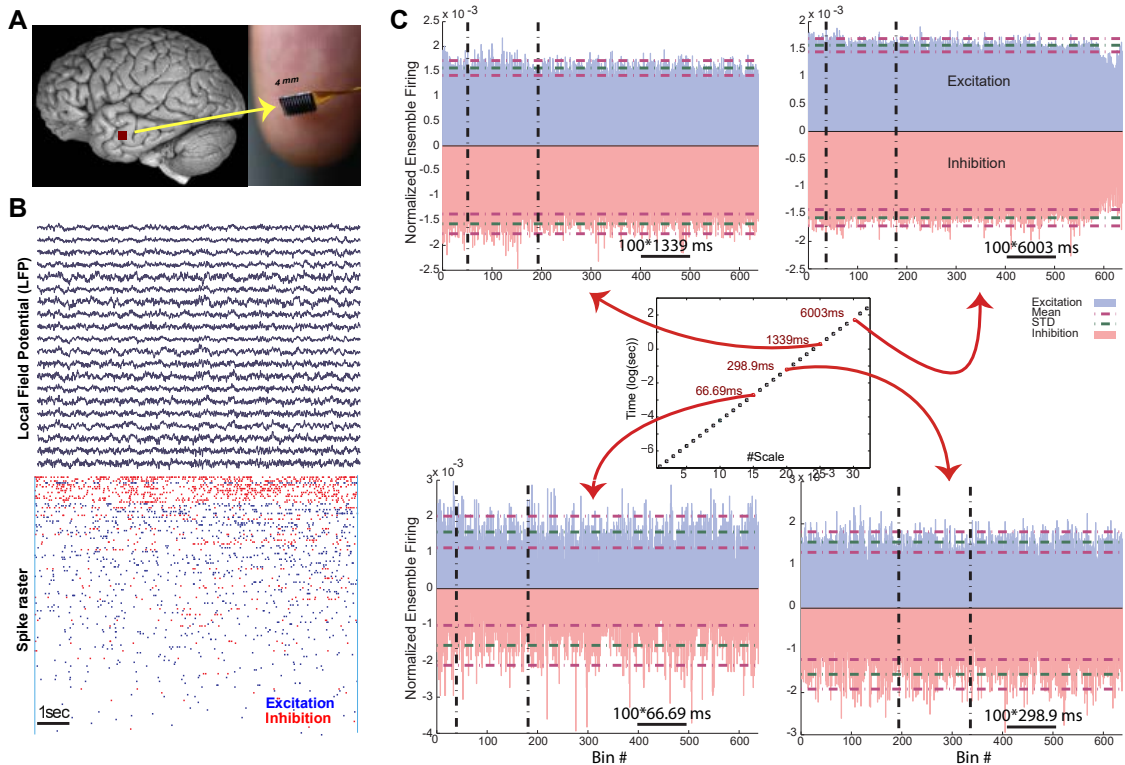}
\caption{
\textbf{Data.} \emph{A.} The implant location (left) and multielectrode array (right). The 10x10 array covers an area with the span of 4mm x 4mm. \emph{B. Sample recording from the multielectrode array.} 
\emph{C.} \textbf{Multiscale representation of balanced Excitation and Inhibition} \emph{Ensemble spiking Excitation-Inhibition balance.} Sample ensemble spiking with different levels of coarse-graining are shown in the four color panels (Note, these panels are example subsets from the rastergram and LFP shown in B. Preservation of Excitation-Inhibition balance across scales shows mirrored activity for different lengths of times (scales). The negative sign is only used conventionally to represent the opposing nature of Inhibition (red) versus Excitation (blue). In each panel, the ensemble firing of a given cell category is normalized by the total firing power in the same category during the shown epoch. Each panel has 656 bins and the duration of each bin is shown in the middle panel. For example, the top right panel represents scale 30 and a duration of 1.0938 hr, showing 656 bin's duration of 6003(sec). The mean (green) and standard deviations (magenta) of each epoch are shown with dashed horizontal lines, respectively. Vertical bars show the time window from which another coarser scale 656-bin epoch was chosen. The middle panel shows the even logarithmic spacing of scales, which was chosen for computational efficiency.
}
\label{fig:Data} 
\end{figure}
\twocolumngrid
\end{widetext}

Here, we build on these perspectives to investigate the \emph{“edge of symmetry”} in cortical dynamics:  
\begin{itemize}
    \item We examine the \textit{scale invariance} of E/I ensemble balance by coarse-graining spiking data from hundreds of excitatory and inhibitory neurons across multiple temporal resolutions.  
    \item We introduce new descriptors, such as \textit{collapse curves}, \textit{partition curves}, and \textit{MCNM (Multiscale Normalized Co-Occurrence Matrix)} to quantify how well E/I fluctuations preserve or lose structure across scales.
    \item We study how \textit{symmetry pathologically breaks down} in pathological seizure states, where excitatory and inhibitory populations become transiently decoupled, potentially reflecting a large-scale breakdown in the underlying balance.
\end{itemize}

The paper is organized as follows. In Section~\ref{sec:methods}, we detail the electrode recordings, classification of excitatory vs. inhibitory units, and the rationale for our coarse-graining approach. In Section~\ref{sec:results}, we present our main findings on E/I scale invariance, randomization tests, and the implications for cortical states from wakefulness to sleep, as well as pathological seizures. Section~\ref{sec:discussion} discusses how these observations relate to brain criticality, multi-scale neural computation, and renormalization ideas, highlighting open directions for future research.

Ultimately, we argue that bridging statistical mechanics with large-scale neuroscience data offers fresh insights into fundamental cortical phenomena. By tracking how E/I ensemble balance persists or fractures across scales, we can potentially uncover new order parameters and emergent laws governing cortical function—laws that may clarify, for instance, why the system hovers near certain critical regimes under normal operation yet can abruptly lose stable balance during pathological events.

\section{Materials and Methods}\label{sec:methods}
\subsection{Data Acquisition and Experimental Procedures}
\label{sec:exp}
Recordings were obtained via multielectrode array implants (\textit{Utah} array, Blackrock Microsystems) in the layers II/III  of human temporal cortex. Each array consists of a 10$\times$10 grid of electrodes with a 400~$\mu m$ inter-electrode spacing, covering an area of roughly 4$\times$4~mm (Fig.\ref{fig:Data}.A). Typically, 4 corner electrodes are reserved for grounding, leaving 96 active channels.

Data were sampled at 30~kHz and stored along with patient video, electrocorticogram (ECoG), and scalp EEG for state labeling (awake, light sleep, slow-wave sleep, REM). The implantation procedure and data collection were performed with informed consent under protocols approved by the Massachusetts General Hospital Institutional Review Board (IRB), in accordance with the Declaration of Helsinki.

\paragraph{Spike Sorting and Unit Classification.}
Each electrode signal was filtered and thresholded to detect potential spikes. Cluster-cutting and morphological criteria were used to identify single-unit activity. We combined short-latency cross-correlogram signs and spike-waveform features to classify units as either excitatory (E) or inhibitory (I) \citep{Bartho2004,Peyrache2012}. 

\subsection{Ensemble Excitation and Inhibition: coarse-graining and renormalization}
\label{sec:ensemble}
Each neuron $n$ can be described by its spiking series $S_n(t)$ over time. To capture the collective dynamics of neuronal populations, we aggregate all excitatory neuron spikes into an ensemble time series, $Ens_E$, and similarly for inhibitory neurons, $Ens_I$. Temporal coarse-graining is performed by summing all spikes within a chosen bin size $\Delta t$, ranging from 1 millisecond to 10 seconds, spaced logarithmically to ensure computational efficiency. The ensemble series are then normalized by the number of excitatory or inhibitory neurons ($N_E$ and $N_I$, respectively):

\begin{equation}
\begin{split}
Ens_E(t;\Delta t) &= \frac{1}{N_E} \sum_{n \in E} \sum_{t \in [t,\,t+\Delta t]}S_n(t), \\
Ens_I(t;\Delta t) &= \frac{1}{N_I} \sum_{n \in I} \sum_{t \in [t,\,t+\Delta t]}S_n(t).
\end{split}
\end{equation}

%
This coarse-graining approach is inspired by renormalization group concepts from statistical physics, which focus on identifying invariant properties across scales \citep{Wilson1971,Wilson1975,Kadanoff1967}. While renormalization traditionally involves rescaling spatial interactions, our method examines invariances in temporal dynamics—specifically, the balance of excitation and inhibition—across diverse timescales. By normalizing spiking data at a population level, this method emphasizes collective cortical behavior and avoids confounding effects of massive subsampling and spatial nonuniformity inherent in high-dimensional neural recordings.
To address subsampling and nonuniformity challenges, we normalized the ensemble series independently for excitatory and inhibitory neuron categories. This adjustment compensates for anatomical and functional biases, such as the 4:1 ratio of excitatory to inhibitory neurons and the 4:1 higher spiking rate of inhibitory neurons observed in anatomical studies \citep{Bartho2004,Peyrache2012}. As a result, the ensemble fractions of excitation and inhibition provide a more robust macroscopic representation of the cortical dynamics.

It is critical to distinguish this \emph{ensemble} E/I balance from the \emph{input} E/I balance described in previous studies \citep{Renart2010,vanVreeswijk1996}. Whereas input E/I balance examines excitatory and inhibitory inputs to individual neurons, ensemble E/I balance focuses on the collective activity of populations, revealing multiscale dynamics that extend beyond local stability, in order to describe macroscopic cortical states.

\subsection{Randomization of Ensemble Spike Times}
\label{sec:random}
To check whether the observed E/I co-fluctuations could be explained by trivial statistical effects or by random alignment of spikes, we created surrogate ensemble time series using two different randomization techniques. Both procedures disrupt the precise temporal matching of excitatory and inhibitory spikes but preserve each category’s overall spike count and distribution.

\subsubsection{Random Permutation of Inter-Spike Intervals (ISI).}
We first compute the pooled ensemble spike train for category $E$ or $I$ and measure its empirical ISI distribution. We then perform a random permutation of those ISIs and reconstruct a \emph{new} spike train by a cumulative sum of the shuffled ISIs \citep{Dehghani2016,Dehghani2012}. Formally,
\begin{equation}
\begin{split}
ISIs_{E,\,\text{shuffled}} \;=\;\text{Permute}\bigl(ISIs_E\bigr),
\\
t^*_{k+1} \;=\; t^*_k \;+\;ISIs_{E,\,\text{shuffled}}[k],
\end{split}
\end{equation}
where $t^*_k$ are the new spike times. The same procedure applies to the inhibitory ensemble $I$. This shuffling preserves the total number of spikes and the distribution of ISIs for each population, but removes any original alignment between $Ens_E$ and $Ens_I$.

The loss of alignment between $Ens_E$ and $Ens_I$ produces fluctuations that lack the tightly bound relationship observed in the real data, underscoring the non-trivial nature of the observed E/I balance in the actual recordings (see results).

\subsubsection{Circular Shift of Each Unit’s Spike Train.}
In a second approach, each unit’s spike train is circularly shifted by a random offset drawn from $\{1,\dots,\max(\text{ISI})\}$ of that unit’s spike series \citep{Dehghani2012}. This effectively disrupts correlations across units in the same ensemble, while preserving internal spike timing of individual neurons. Again, we aggregate the circular-shifted neurons into a randomized $Ens_E$ or $Ens_I$. 

The comparison of this method to the real data highlights how maintaining internal timing within individual neurons is insufficient to reproduce the observed E/I dynamics. Instead, the results suggest that the balance relies critically on synchronized interactions between neurons across the population.

Both methods yield surrogates $Ens_E^{(\text{rand})}$, $Ens_I^{(\text{rand})}$ that can then be coarse-grained at each $\Delta t$ and compared to the real data. Conceptually, the first method more drastically alters the internal structure of the entire ensemble spike train, whereas the second method maintains each unit’s local spike times but randomizes inter-unit alignment. 
In both cases, the loss of E/I balance is evident, as demonstrated by the scattering of points in the $Ens_E$--$Ens_I$ space, which contrasts with the tightly bound line observed in the real data (see results).

\subsection{Multi-scale Analysis: Collapse Curves}
\label{sec:collapse}
Following ideas from crackling noise and avalanche theory \citep{Papanikolaou2011,Sethna2001,Friedman2012}, we define 
\[
f_{\text{obs}}(t)\;=\;Ens_E(t) \;-\; Ens_I(t)
\]
as a fluctuation measure of ensemble E--I balance. We sort the values of $f_{\text{obs}}$ in ascending order, then perform a partial sum or cumulative distribution. Plotting these sums, when properly rescaled, can reveal whether the curves at different $\Delta t$ “collapse” onto a universal shape \citep{Sethna1992,Sethna2006}. Formally, for each timescale $\Delta t$:
\begin{equation}
F_{\Delta t}(\alpha) \;=\; \sum_{\{t: f_{\text{obs}}(t) \le \alpha\}} f_{\text{obs}}(t),
\qquad
\alpha \in \mathbb{R}.
\end{equation}
We then rescale $\alpha$ and $F_{\Delta t}(\alpha)$ to lie in $[0,1]$; a perfect overlap of $F_{\Delta t}$ for multiple $\Delta t$ suggests scale invariance of the distribution of $f_{\text{obs}}$. 

In the context of cortical networks, such scale-invariant behavior may reflect underlying mechanisms of self-organization. The collapse curves serve as a diagnostic tool, revealing whether the observed E/I fluctuations exhibit universal scaling properties, distinct from those of random surrogates. For instance, deviations in collapse curves can highlight state transitions, such as those between wakefulness and sleep, or point to pathological states where the E/I balance breaks down.

We compare the real data’s collapse curves to those of the random surrogates. If randomization “whitens” the distribution, destroying correlations between $Ens_E$ and $Ens_I$, the surrogate collapse curves do not match those of the original data \citep{Dehghani2016,Dehghani2012}. Distances in collapse-curve space can be used to quantify the degree of self-similarity or scale invariance.

By quantifying differences between real and surrogate collapse curves, we can systematically investigate the degree of scale invariance and identify signatures of neural coordination across states and scales.

\subsection{Partition Curve for Ensemble Activity}
\label{sec:partition}
To assess the distribution of ensemble firing rates more explicitly, we use a \emph{partition curve}, which plots the cumulative percentage of total activity contributed by segments of ensemble activity of increasing size. Let $A$ denote a random variable representing the fraction of ensemble activity within a given functional category (E or I), where $A$ takes values $b_1 \leq b_2 \leq \dots \leq b_n$ with associated probabilities $g(b_j) = \Pr(A = b_j)$. The cumulative sums are defined as:
\begin{equation}
C_i = \sum_{j=1}^i g(b_j),
\quad
D_i = \sum_{j=1}^i g(b_j)b_j,
\quad
P_i = \frac{D_i}{D_n},
\end{equation}
where $C_i$ represents the cumulative proportion of activity up to the $i$-th segment, $D_i$ is the weighted sum of activity values up to that segment, and $P_i$ normalizes the cumulative activity by the total activity $D_n$.
Plotting $(C_i, P_i)$ in a $[0,1]\times [0,1]$ coordinate system yields a convex curve that lies below the diagonal line $y=x$, which represents a perfectly uniform distribution of activity. The distance of the partition curve from the diagonal quantifies the \emph{unequivalence} of ensemble activity. A curve closer to the diagonal indicates more equal distribution of activity across segments, while a curve further away reflects greater disparities in the distribution.
By constructing partition curves at various timescales $\Delta t$, we examine whether the distribution of excitatory or inhibitory ensemble activity remains consistent across scales or whether coarser timescales homogenize the firing distribution. These curves reveal differences between the real data and surrogate randomizations, with surrogates typically producing ``steeper (less convex)'' curves indicative of more uniform distributions.

\subsection{Multiscale Normalized Co-occurrence Matrix (MNCM)}
\label{sec:co-occurrence}
To analyze the properties of the joint probability of the fraction of Excitation and Inhibition across many temporal scale, we implemented a multidimensional textural feature analysis. From the three-dimensional space of excitation/inhibition/scale (EIS), we calculated a \emph{Multiscale Normalized Co-occurrence Matrix} (\emph{MNCM}) following volumetric adaptations of textural analysis techniques \citep{Haralick1973,Davis1979,Tesa2008}. The \emph{MNCM} quantifies the frequency of co-occurrence of pairs of discrete normalized values $(i,j)$ (representing excitation and inhibition levels, respectively) across multiple spatial relationships or offsets within the EIS space.
Each entry in the \emph{MNCM} is given by the normalized co-occurrence probability:
\begin{equation}
p(i,j);=;\frac{P(i,j)}{\sum_{i,j} P(i,j)},
\end{equation}
where $P(i,j)$ represents the number of times excitation level $i$ co-occurs with inhibition level $j$ for a specific offset. The normalization ensures that $\sum_{i,j} p(i,j) = 1$. $N_g$ denotes the number of distinct normalized levels in the EIS space.

\subsubsection{Feature Categories Derived from MNCM}
To extract meaningful insights from the \emph{MNCM}, we computed features across three categories:

\begin{enumerate}
    \item \textbf{Descriptive Statistics Group}:
This category includes metrics that describe the overall distribution of excitation and inhibition co-occurrence in the EIS space.

\textit{Correlation}: Measures the linear dependency between excitation and inhibition levels:
\begin{equation}
Correlation = \cfrac{\sum_i \sum_j (i,j)p(i,j)-u_x u_y}{\sigma_x \sigma_y},
\end{equation}
where ($u_x$) and ($u_y$) are the means of the marginal distributions ($p_x(i)$) (probability of excitation level $i$) and ($p_y(j)$) (probability of inhibition level $j$), respectively, and ($\sigma_x$) and ($\sigma_y$) are their standard deviations.

\item \textbf{Contrast Group}:
This category captures variations or contrasts in the co-occurrence of excitation and inhibition:
\begin{itemize}
    \item \textit{Contrast}: Reflects the intensity of local variations in the \emph{MNCM}:
    \begin{equation}
Contrast = \sum_{i=1}^{N_g} \sum_{j=1}^{N_g} (i-j)^2 p(i,j) 
\end{equation}
    \item \textit{Dissimilarity}: Similar to contrast but with linear weights, measuring deviations between excitation and inhibition:
\begin{equation}
Dissimilarity = \sum_{i=1}^{N_g} \sum_{j=1}^{N_g} (i-j) p(i,j) 
\end{equation}
    \item \textit{Homogeneity}: Indicates the closeness of the co-occurrence distribution.
\begin{equation}
Homogeneity = \sum_{i=1}^{N_g-1} \sum_{i=1}^{N_g-1} \cfrac{1}{1+(i-j)^2}p(i,j)
\end{equation}
\end{itemize}

\item \textbf{Orderliness Group}:
This category reflects the regularity or predictability of excitation-inhibition co-occurrence patterns:
\begin{itemize}
    \item \textit{Energy}: Represents the sum of squared values in the \emph{MNCM}, quantifying uniformity in co-occurrence:
\begin{equation}
Energy =  \sum_{i=1}^{N_g}  \sum_{j=1}^{N_g} p(i,j)^2  \equiv sqrt(\:Angular \:Second \:Moment)  
\end{equation}
    \item \textit{Entropy}: Measures the degree of disorder in the co-occurrence matrix:
\begin{equation}
Entropy =  -\sum_{i=1}^{N_g} \sum_{j=1}^{N_g} p(i,j)log(p(i,j))
\end{equation}
\end{itemize}

\end{enumerate}

\subsubsection{Interpretation of MNCM Features}
Each feature group captures a distinct aspect of excitation-inhibition interactions:

\begin{itemize}
    \item The \textit{Descriptive Statistics Group} highlights the overall co-occurrence structure and the relationship between excitation and inhibition levels, with correlation being a key metric for dependency.
    \item The \textit{Contrast Group} measures local variations in co-occurrence, with homogeneity indicating how similar excitation and inhibition levels are.
    \item The \textit{Orderliness Group} focuses on the balance of regularity and complexity, with entropy quantifying  disorder at a given scale.
\end{itemize}
By computing these features across multiple temporal scales independently and aggregating them, we generate a comprehensive feature vector that captures the multiscale dynamics of excitation and inhibition interactions. Comparing these feature vectors across wake, sleep, and seizure states allows us to detect patterns of symmetry breaking or scale-specific disruptions in E/I balance \citep{Dehghani2016,Dehghani2012}.

\subsection{Synaptic Current, Wavelet Coherence, and Seizure Analysis}  
\subsubsection{Synaptic Current}
\label{sec:syncur}  
To study the multiscale dynamics of spiking excitation and inhibition alongside the multiscale representation of local field potentials (LFPs), we reconstructed the estimated synaptic current from unit activity. Synaptic current is widely regarded as a primary generator of LFPs \citep{Niedermeyer2005,Nunez2006}, reflecting the collective activity of tens of thousands of neurons within the recording area where spikes are sample. The modeling process involved convolving each spike with an exponential kernel \citep{Bedard2006}:  
\begin{equation}\label{eq:synaptic_current}  
C(t) = \int_{-\infty }^{\infty } D({t}')\exp\left[-\frac{t-{t}'}{\tau_{s}}\right]d{t}',  
\end{equation}  
where:  
\begin{itemize}  
    \item $C(t)$ represents the estimated synaptic current,  
    \item $D(t')$ is the spike train,  
    \item $\tau_{s}$ is the synaptic time constant.
\end{itemize}  

At any given time, the synaptic current represents a weighted sum (or integral, in the continuous case) of the spike train, where the weights decay exponentially as the time difference between the spike and the current time increases. After convolving the spike trains with their respective kernels, the estimated synaptic current was derived by subtracting the inhibitory ensemble current from the excitatory ensemble current. Note that the result was normalized by the number of neurons in each category (excitatory or inhibitory) to correct for sampling biases and ensure a population-level representation of the dynamics \citep{Dehghani2016}.  

For our specific implementation, we computed the net synaptic current as:
\begin{equation}  
C(t) \;=\;\Bigl(\sum_{t'_i \in D_E} e^{-(t-t'_i)/\tau_{\!E}}\Bigr)  
\;-\;\Bigl(\sum_{t'_j \in D_I} e^{-(t-t'_j)/\tau_{\!I}}\Bigr),  
\end{equation}  
where $D_E$ and $D_I$ are the excitatory and inhibitory spike trains, $t'_i$ and $t'_j$ represent individual spike times, and $\tau_{\!E} \approx 3$~ms and $\tau_{\!I} \approx 10$~ms are typical time constants for excitatory and inhibitory post-synaptic currents, respectively \citep{Rudolph2007,Bedard2006_b}.    

\subsubsection{Wavelet Coherence and Ridge Detection}
To relate large-scale oscillations in local field potentials (LFPs) to excitatory and inhibitory spikes, we computed the wavelet coherence of the LFP signal \citep{Torrence1998} and compared it to the estimated synaptic current. Wavelet coherence quantifies the time-frequency relationship between these signals and measures localized phase-locked activity based on the wavelet cross-spectrum between two signals.

The wavelet cross-spectrum is defined as the product of the wavelet transform of signal $(x)$ and the complex conjugate of the wavelet transform of signal $(y)$:

\begin{equation}
W_{n}^{XY}(s) = W_{n}^{X}(s)W_{n}^{Y^{*}}(s) 
\end{equation}
where $W_{n}{Y}(s)$ are the wavelet transforms of signals $x$ and $y$, respectively, at scale $s$, and $*$ represents the complex conjugate.
Wavelet coherence is then calculated as the squared absolute value of the smoothed cross-spectrum, normalized by the product of the smoothed individual power spectra:

\begin{equation}
C_{n}^{2}(s) = \frac{\left | \left \langle  W_{n}^{XY}(s)\right \rangle.s^{-1}\right |^{2}} 
{\left \langle \left |W_{n}^{XX}(s) \right | .s^{-1} \right \rangle 
 \left \langle \left |W_{n}^{YY}(s) \right | .s^{-1} \right \rangle} 
\end{equation}
where $\langle \cdot \rangle$ denotes smoothing in both time and scale. The result, $C_{n}^{2}(s)$, is a measure of the coherence between signals $x$ and $y$ as a function of time and frequency, ranging from 0 (no coherence) to 1 (perfect coherence).

To identify ridges in the wavelet coherence, we used the \emph{crazy-climber algorithm}, a stochastic relaxation method based on Markov chain Monte Carlo (MCMC) \citep{Carmona1997,Carmona1999}. This method extracts multiple ridges from the time-frequency representation by iteratively locating regions of high energy. The algorithm operates by sampling the energetic distribution in the time-frequency plane and identifying paths of maximum coherence across different frequency bands.

Wavelet ridges correspond to regions of strong coherence, enabling us to track dominant phase-locked interactions between the LFP and synaptic current. These ridges capture frequency ranges where excitatory and inhibitory dynamics are most tightly synchronized. Rapid transitions or disruptions in ridge patterns often coincide with significant deviations in synaptic current ($C(t)$), particularly during seizure onset \citep{Franaszczuk2011,Zaveri2011,Schevon2012}.

By combining wavelet coherence and ridge detection, we characterize the temporal and spectral dynamics of LFPs and link them to the underlying multiscale balance of excitation and inhibition. This approach reveals how seizures can emerge from large-scale breakdowns in E/I synchronization, emphasizing the critical role of symmetry breaking in cortical dynamics.

\section{Results $\&$ Discussion}\label{sec:results}

\paragraph{Temporal coarse-graining framework.} 
Cortical neural activity operates across multiple spatial and temporal scales, presenting unique challenges for analysis. To address this, we employ a \emph{temporal coarse-graining} procedure inspired by renormalization group concepts from statistical physics where system properties transcend many different length scales \citep{Kadanoff1967,Wilson1971, Wilson1979}. While our approach differs from traditional spatial renormalization, it serves a similar purpose by identifying invariant properties across temporal scales.

As described in Methods, we aggregate excitatory and inhibitory spiking activities into respective ensemble time series and normalize by neuron count. The observable is the difference signal:

$f_{\mathrm{obs}}(t;\Delta t)=\mathrm{Ens}_E(t;\Delta t)-\mathrm{Ens}_I(t;\Delta t)$

By varying the bin width $\Delta t$ from 1 ms to 10 s, we examine whether this population-level observable maintains structural invariance under successive coarse-grainings —-the hallmark of a renormalization group fixed point. In the temporal domain, this would correspond to a diverging correlation \emph{time}, analogous to correlation length in spatial renormalization \citep{Samorodnitsky2016,Garcia-Perez2018}.

This approach offers two significant advantages. First, it renders our analysis agnostic to single-neuron spiking variability, focusing instead on collective dynamics. Second, it addresses challenges of subsampling and spatial nonuniformity inherent in high-dimensional neural recordings. Below, we quantify the structural properties of ensemble excitation-inhibition balance and demonstrate how these properties persist across various brain states and timescales.

\subsection{Multiscale Signatures of Balance Invariance}
    \subsubsection{Excitation–Inhibition Balance Across Temporal Scales}
        \paragraph{Scale-invariant mirroring and variance scaling.} 
        Across all functional states—wake, REM, light sleep (LS), and slow-wave sleep (SWS)—ensemble excitation and inhibition consistently track one another (Fig.\,\ref{fig:Data}B,C). When normalized by total firing power, $\mathrm{Ens}_{E}$ and $\mathrm{Ens}_{I}$ exhibit scaled mirroring of each other across temporal scales from 1 ms to 10 s. This invariance suggests a systematic preservation of balance that transcends specific brain states.
        
To investigate whether this balance results from simple averaging, we analyzed the variance scaling of the difference signal $f_{\mathrm{obs}}(t;\Delta t)$. If excitatory and inhibitory ensembles were independent, the central limit theorem would predict variance to decrease linearly with bin size ($\propto 1/\Delta t$). Our analysis reveals a sublinear scaling relationship, with variance decreasing more slowly than predicted by simple statistical averaging. This departure from central-limit behavior indicates structured correlations between excitation and inhibition that persist across scales.

When examining $\mathrm{Ens}_{E}$ versus $\mathrm{Ens}_{I}$ distributions, the data form elongated ellipsoids with major axes aligned along the diagonal of perfect balance (Fig.\,\ref{fig:NoiseCollapse}A1). While the width orthogonal to this diagonal—representing instantaneous dominance of either excitation or inhibition—contracts with increasing $\Delta t$, the orientation and symmetry remain preserved, further demonstrating the scale-invariant structure of E-I balance.

        \paragraph{Dispersion metrics across states.}
        Quantitative analysis reveals remarkably consistent dispersion indices across all behavioral states (Table~\ref{tab:Table1}). Mean absolute deviation (MAD) remains below 0.003, indicating minimal overall variability. Skewness values stay near zero ($|\gamma|<0.055$), confirming symmetrical fluctuations around perfect balance. The coefficient of variation (CV = $\sigma/\mu$) for excitatory and inhibitory ensembles exhibits a proportional relationship across scales, with the ratio between them maintaining a near-unity slope ($S_{CV}\approx1$) for all vigilance states. This proportional scaling of variability provides additional evidence that excitatory and inhibitory activities remain tightly coupled even as their absolute fluctuations decrease with temporal coarse-graining.
\begin{table}[h]
\caption{\label{tab:Table1} Indices of dispersion. The Mean Absolute Deviation (MAD) indicates overall variability, Skewness measures the symmetry of the distribution, and the Slope of the Coefficient of Variation ($S_{CV}$) reflects the stability of variability across scales. These measures demonstrate the consistency of the E/I balance across different temporal scales.}
\begin{tabular}{ccc|c}
 & MAD\footnotemark[1]& Skewness\footnotemark[2] & $S_{CV}$\footnotemark[3]\footnotemark[4]\\
\hline
AWAKE & \(0.0023\pm0.0030\) & \(-0.013\pm0.046\) & \\
REM & \(0.0028\pm0.0032\) & \(0.052\pm0.015\) &   $|S_{CV} - 1| < \epsilon $\\
LS & \(0.0021\pm0.0024\ \) & \(-0.010\pm0.015\) &  $\epsilon < 10^{-2}$\\
SWS & \(0.0027\pm0.0034\) & \(-0.012\pm0.087\) &  \\
\end{tabular}
\end{table}
\footnotetext[1]{Mean absolute value deviation, \(\frac{1}{n}\sum_{i=1}^n|x_i-\overline{X}|\)}
\footnotetext[2]{\(\gamma = \operatorname{E}\Big[\big(\tfrac{X-\mu}{\sigma}\big)^{\!3}\, \Big]\)}
\footnotetext[3]{$CV = {\sigma}/{\mu}$}
\footnotetext[4]{$S_{CV}$:Slope of multiscale E:I Coefficient of Variation. Various states returned similar value $S_{CV}$.}

    \subsubsection{Randomization Tests: Disrupting Temporal and Inter-Unit Structure}
    To verify that the observed scale-free balance depends on precise spike timing, we generated two types of surrogate data that preserve marginal statistics while disrupting different aspects of correlation structure:

\paragraph{Random permutation of inter-spike intervals.}
For each functional pool (E or I), we computed the empirical inter-spike interval (ISI) distribution from the pooled ensemble series, randomly permuted these intervals, and reconstructed a new spike train through cumulative summation. This method preserves the total spike count and ISI distribution while disrupting the temporal order of aggregate spikes. Conceptually, this procedure acts as a linear whitening transformation by removing temporal correlations in the ensemble spike trains.

\paragraph{Circular shift of individual spike trains.}
In this approach, each unit's spike train was circularly shifted by a random offset drawn uniformly from $[1,\max\text{ISI}_{\!n}]$. This method preserves the internal spike timing structure within individual neurons but eliminates cross-neuron temporal alignment. By applying the transformation independently to each neuron, this procedure implements a non-linear whitening: it removes higher-order dependencies between units while maintaining non-Gaussian marginal distributions.

  \begin{widetext}
  \onecolumngrid
\begin{figure}[tbh]
\includegraphics[scale=0.78]
{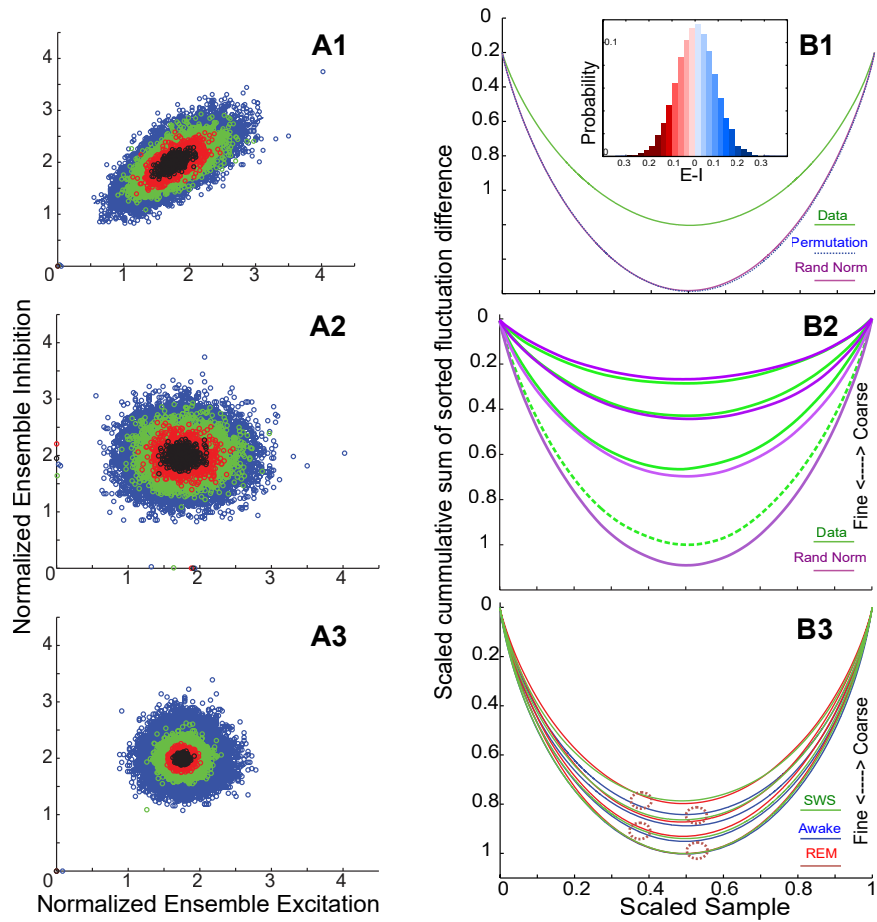}
\caption{
\textbf{Randomization of ensemble activity.} \emph{A1: Data.} Scatterplot of normalized ensemble excitation and inhibition. Blue, red, green and black represent 4 sample scales. Paired E-I scatter as elipsoids with their major axis of variation along the diagonal (perfect symmetry). \emph{A2:Random permutation of ISI in the ensemble series.} \emph{A3:Fixed-ISI circular shift of spikes.} Ensemble Excitation and Inhibition show correlated and balanced structure that is disrupted by sphering randomization. For details of the randomization techniques, see methods. 
\textbf{Fluctuation collapse.} \emph{B1. Fine-scale collapse curve versus random surrogates.} The inset shows the histogram distribution of fluctuation of ensemble activity at the finest scale. \emph{B2. Comparison of collapse curves across different scales.} Note that at coarsest scale, surrogate collapse curve approaches the data collapse curve. Different random surrogates collapse onto the same curve but distinctively are different from data collapse curve. \emph{B3. Comparison of collapse curves for different states (SWS: slow-wave sleep, REM: Rapid Eye Movement, and Awake) across various scales.} Dashed circles are visual guides to exemplify that more or less collapse curves of different states have similar behavior in a given scale.
}
\label{fig:NoiseCollapse} 
\end{figure}
  \twocolumngrid
  \end{widetext}

\paragraph{Effects on E-I structure.}
Both randomization methods dramatically transform the tightly bound, diagonal-aligned ellipsoids observed in the empirical $\mathrm{Ens}_{E}\times\mathrm{Ens}_{I}$ scatter plots (Fig.\,\ref{fig:NoiseCollapse}A1) into circular, isotropic clouds (Fig.\,\ref{fig:NoiseCollapse}A2,A3). This "sphering" effect demonstrates the loss of structured correlation between excitatory and inhibitory populations. Moreover, the variance collapse and coefficient-of-variation scaling observed in the original data disappear in the randomized surrogates, confirming that scale invariance relies on precisely synchronized temporal organization across the examined neural population.

These randomization results underscore a critical insight: balanced E-I dynamics are not simply an inevitable consequence of matching mean firing rates between populations. Rather, they depend fundamentally on structured correlations—both within and between neurons—across timescales from milliseconds to seconds. Consequently, models or analyses that neglect these correlations will fail to capture the essential organizing principle revealed by our multiscale approach.

    \subsubsection{Subsampling Limitations}
    It is important to acknowledge the limitations of current recording technologies. A Utah array samples approximately $10^{2}$ units from the estimated $\sim\!10^{4}$ to $2\!\times\!10^{4}$ neurons within a cortical column \citep{Rakic_1988,Herculano_Houzel_2008}. Our results therefore provide a conservative estimate of balance invariance. As recording technologies advance, they will enable the exploration of excitatory-inhibitory balance with higher spatial resolution across many neighboring columns. Given the underlying functional anisotropy at different spatial scales—such as our prior observation of stereotypical directionality of local field potential traveling waves at the scale of a cortical column \citep{Le_Van_Quyen_2016} and in theoretical models of propagating waves across larger patches of cortex \citep{Galinsky_2020}—the investigation of ensemble excitatory-inhibitory translational symmetry will become increasingly feasible. This exploration will guide us in developing a formalism that could link micro and macroscales.
    
    \subsubsection{Functional Implications and Network Mechanisms}
These randomization results, together with the scale-free properties observed in the original data, point to profound functional implications. The symmetric, scale-free nature of E-I balance significantly expands the dynamical phase-space available for neural computation: inputs can modulate activity over several orders of magnitude without driving the network away from equilibrium \citep{Xing1996}. To understand the mechanistic basis of this non-random scale invariance, we previously compared our findings with a conductance-based balanced network model (COBA) that exhibits self-generated balanced activity \citep{Dehghani2016}. The COBA model demonstrates preserved balance across scales with similar instantaneous deviations from perfect equilibrium as observed in our experimental data. Importantly, this model shows that scaled balance emerges from recurrent interactions within the network rather than from external inputs or simple averaging processes. The sublinear variance scaling we observe in vivo is consistent with this model's predictions, suggesting that the multiscale balance reflects intrinsic network dynamics rather than statistical artifacts.

These findings align with earlier intracellular studies demonstrating precise tuning of spike timing by coordinated excitatory-inhibitory conductances \citep{Higley2006,Rudolph2007}, and with simulations showing that the spatial distribution of inhibitory synaptic activity on pyramidal cells maintains the system in a state of inhibition-dominant excitability, poised for thalamic input \citep{Trevelyan2005}. Collectively, our analyses indicate that balance is enforced by recurrent architecture in a manner that preserves its structure from milliseconds to seconds, providing a robust foundation for cortical computation across diverse timescales and brain states.

\subsection{Universality of Balance}
\subsubsection{Collapse-Curve Analysis of Excitation-Inhibition Fluctuations}

\paragraph{Construction and significance.}
To investigate whether the ensemble excitation-inhibition fluctuations $f_{\mathrm{obs}}(t;\Delta t)=\mathrm{Ens}_{E}-\mathrm{Ens}_{I}$ follow a universal scaling law, we developed the \textit{collapse-curve} technique inspired by methods established for analyzing crackling noise and avalanches in physical systems \citep{Sethna2001,Papanikolaou2011}. This approach involves rescaling different curves of a physical quantity measured at various control parameter values (in our case, timescales) to make them collapse onto a single curve, revealing universality or self-similarity. After sorting $f_{\mathrm{obs}}$ values in ascending order and calculating their cumulative sum, a symmetric parabolic curve emerges. We rescaled each curve for different time scales so that it would be bounded between 0 and 1 on the x-axis and then normalized the y-axis by the product of the number of bins. This normalization ensures a fair comparison of timeseries of varied lengths. When properly rescaled, these curves from different temporal scales should collapse onto a single master curve if the underlying dynamics exhibit scale invariance.

\paragraph{Data versus surrogates.}
The empirical collapse curves from 1 ms to 10 s indeed overlay almost perfectly, forming a symmetric parabola (Fig.\,\ref{fig:NoiseCollapse}B1). In contrast, both surrogate ensembles—generated through ISI permutation and circular shift methods—fail to collapse onto this universal curve. The surrogate curves deviate significantly from the empirical envelope, except at the coarsest temporal scale where fine temporal correlations are largely averaged out (Fig.\,\ref{fig:NoiseCollapse}B2). This finding demonstrates that the scale-free organization of excitation-inhibition balance emerges from precise temporal coordination, rather than being a mere statistical byproduct of matched ensemble firing rates of E and I cell classes.

\paragraph{State invariance.}
When comparing collapse curves across different brain states—wake, REM, light sleep (LS), and slow-wave sleep (SWS)—we observe remarkably similar master curves (Fig.\,\ref{fig:NoiseCollapse}B3). This state invariance further reinforces the renormalization-group concept that microscopic details of different phases become irrelevant once variables are properly coarse-grained, revealing universal properties that transcend specific behavioral states.

\paragraph{Relation to critical phenomena.}
The collapse of properly rescaled observables onto a universal curve represents a defining empirical signature of universality classes near critical points \citep{Friedman2012}. While neural avalanches exhibit similar scaling properties, their interpretation remains debated. Traditional self-organized criticality (SOC) models have proven inadequate in capturing the role of inhibition and the diversity of spatiotemporal patterns observed in neuronal networks \citep{Dehghani2012}. Similarly, maximum entropy approaches fail to accurately represent excitation-inhibition asymmetries \citep{Zanoci2019}. 

Our collapse-curve analysis points instead to a distinct form of criticality in cortical networks, potentially characterized by scale-invariance across multiple parameters. This may arise from interconnected feedback mechanisms—including recurrent excitation, local inhibition, and neuromodulatory gain control—that drive the network toward a manifold of quasi-critical fixed points rather than a single SOC attractor. The scaling properties we observe suggest that the cortex operates in a special region of phase space: poised near criticality yet precisely regulated by inhibitory feedback mechanisms.

In summary, ensemble excitation-inhibition fluctuations follow a universal scaling form that persists across four orders of magnitude in time and through all vigilance states, yet disintegrates under randomization. This universality suggests that balanced cortical activity represents a fundamental organizing principle of neural computation, rather than a coincidental statistical property.

\subsubsection{Partition-Curve Analysis: Distribution of Ensemble Firing}

\paragraph{Limitations of traditional spiking models.}
Classical Poisson models, which treat neuronal firing as a continuous-time Markov process with exponentially distributed inter-spike intervals \citep{Shadlen1994,Rieke1999}, fail to accurately capture realistic neuronal dynamics. This inadequacy stems primarily from their inability to account for neuronal refractoriness \citep{Berry2200,Amarasingham2006,Deger2012}, \textit{in-vivo} high-conductance state variability \citep{Telenczuk2017_b}, and serial dependencies that violate renewal theory assumptions \citep{Gerhard2017,Gerstner2014}. Renewal theory extends the Poisson process with arbitrary holding times but still assumes independent, identically distributed intervals—an assumption violated in neurons with strong adaptation, where interspike intervals show complex dependencies \citep{Chow1996,Mainen1995}. While more sophisticated formalisms exist—including time-dependent renewal theory and Hawkes processes—these become unwieldy when applied to strongly non-stationary (neural population) data \citep{Hodara2017}. To overcome these limitations, we introduce a distribution-free approach using partition curves to analyze ensemble activity.

  \begin{widetext}
  \onecolumngrid
\begin{figure}[tp]
\includegraphics[scale=0.45]
{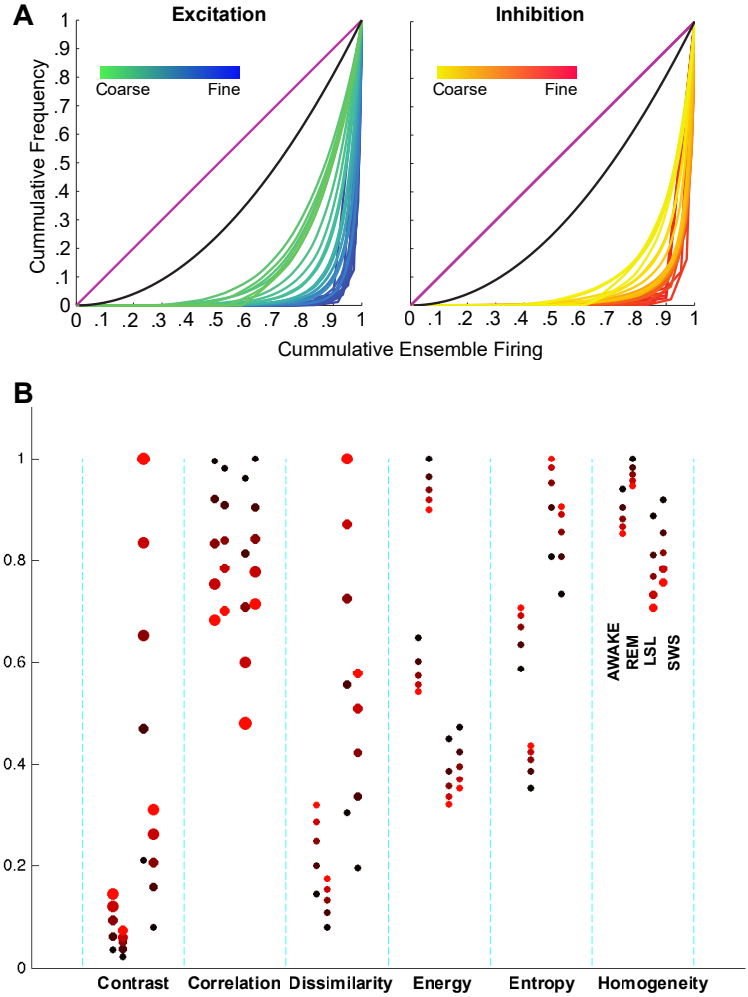}
\caption{
\textbf{A:Ensemble activity partition curve.} In each panel, partition curves for various scales are displayed against the diagonal (magenta) line of equivalence. A perfectly equal ensemble size distribution would be one in which every ensemble size has the same frequency (magenta line). The black curve represents the partition curve for the normal distribution surrogate series. Partition curves for different scales are color-coded, with brighter colors indicating coarser scales and darker colors indicating finer scales. With increasing coarse-graining, the partition curves approach closer to the diagonal and random partition curves but always maintain the characteristic non-equivalence property. 
\textbf{B:Multidimensional features of Excitation and Inhibition co-occurrence across different states.} In each vertical band, a distinct feature (extracted from the multidimensional co-occurrence matrix of Excitation/Inhibition/Scale) is shown for various states; Awake, REM (rapid-eye movement), LSL (light sleep) and SWS (slow-wave sleep) from left to right. The features belong to three main categories:
\emph{contrast group} (contrast, homogeneity and dissimilarity), \emph{descriptive group} (correlation), and \emph{orderliness group} (entropy and energy). Colors brighten as the distance of the matched pairs in the co-occurrence matrix increases. The radius of each circle shows the std. The features define the state-dependent properties of Excitation and Inhibition interaction in a multiscale computational framework. Note that each category of measures is normalized to the maximum across different scales.
}
\label{fig:LorenzHarallick} 
\end{figure}
  \twocolumngrid
  \end{widetext}

\paragraph{Partition curve methodology.}
The partition curve provides a graphical representation of ensemble spike firing distribution without making distributional assumptions. For a given timescale $\Delta t$, we bin ensemble spike counts, rank-order them $\{b_1 \leq b_2 \leq ... \leq b_n\}$, and plot the cumulative fraction of bins $C_i = \sum_{j \leq i}1/n$ against the cumulative fraction of spikes $P_i = \sum_{j \leq i}b_j/\sum_k b_k$. This approach is analogous to a probability-probability plot comparing the distribution against a hypothetical uniform distribution. The diagonal line represents perfect equivalence of normalized ensemble activity across time, while increased deviation from this line indicates higher degree of unequivalence in the distribution. The partition curve's utility extends to nonstationary data, allowing for temporal distribution comparisons not easily achieved with conventional methods.

\paragraph{Scale-dependent firing inequality.}
As shown in Figure \ref{fig:LorenzHarallick}A, the partition curves are consistently convex and clearly separated from both the diagonal and a Gaussian surrogate (black curve), demonstrating that ensemble activity is highly inhomogeneous. Color-coded curves representing different timescales (darker colors for finer scales, brighter for coarser) reveal that as coarse-graining increases, partition curves approach the diagonal, reflecting partial homogenization of firing rates. However, a significant ``inequality gap'' persists even at the coarsest scale, indicating that cortical activity maintains burst-like structure across all examined timescales. 

\paragraph{Functional implications.}
The persistent inequality gap quantifies the volume of functionally accessible microstates under specific coarse-graining conditions. This finding aligns with evidence that strong adaptation and network interactions generate serial inter-spike interval correlations that extend beyond simple renewal theory \citep{Bryant1976,Chow1996,Mainen1995}. The partition-curve analysis complements collapse-curve results: while excitatory and inhibitory ensembles maintain balance on average, their constituent spikes are unevenly distributed in time, creating a scale-invariant balance substrate upon which dynamics operate.

\subsection{Symmetry Breaking}
\subsubsection{Wake-Sleep Modulations Captured by Multiscale Co-occurrence Features}

\paragraph{Conceptual foundation in renormalization.}
As we transition toward coarser temporal scales, the joint distribution of excitatory and inhibitory activity tends to simplify, reflecting how macroscopic behaviors often exhibit more universal properties than microscopic ones. This simplification is conceptually linked to renormalization group (RG) theory, where a ``fixed point'' represents a state of invariance under successive transformations, providing a crucial test for criticality beyond just thermodynamic analogies \citep{Meshulam_2019}. While our coarse-graining approach successfully integrates out finer-scale degrees of freedom, the emergence of distinct macroscopic states suggests that the system undergoes symmetry breaking—deviations from perfect excitation-inhibition balance that characterize different functional cortical states.

To quantify these state-dependent deviations, we introduce a multidimensional feature analysis in the three-dimensional space of Excitation:Inhibition:Scale (EIS). We computed a MNCM following techniques adapted from volumetric texture analysis \citep{Haralick1973,Davis1979,Tesa2008}. This approach quantifies how frequently specific excitation and inhibition levels co-occur across different spatial relationships within the EIS space. In this context, ``distance'' refers to an offset parameter that defines the relationship between pairs of points in the EIS space—these points represent joint occurrences of specific E and I values across scales, not physical locations. Each MNCM entry indicates the frequency at which a particular E value co-occurs with a particular I value at a given offset distance.

From these co-occurrence matrices, we extracted features in three key categories:
\begin{itemize}
    \item \textbf{Descriptive Statistics} — measures like mean, variance and correlation that characterize the overall relationship between excitation and inhibition values.
    \item \textbf{Contrast Measures} — metrics such as contrast, homogeneity, and dissimilarity that capture local variations in excitation-inhibition patterns.
    \item \textbf{Orderliness Measures} — features like energy (angular second moment) and entropy that quantify the regularity or complexity of the co-occurrence distribution.
\end{itemize}

\paragraph{State-dependent signatures across vigilance states.}
Figure \ref{fig:LorenzHarallick}B presents six normalized key features across awake, REM, light sleep (LS), and slow-wave sleep (SWS) states, revealing distinctive patterns of E-I coordination symmetry breaking:

\textbf{Energy} (a measure of uniformity) shows highest values in awake and REM states, with a notable distinction between these two states. Its significantly lower values in light and slow-wave sleep may reflect the emergence of slow oscillatory activity in deeper sleep states. The distinctive elevation of energy in REM compared to wakefulness could be related to the unique internally-driven information processing requirements of dream states.

\textbf{Entropy} displays an inverse pattern, peaking in SWS and LS while reaching minimal levels in REM. This suggests different degrees of variability in E-I patterns across states, potentially related to changes in cortical-thalamic coupling during different phases of the sleep-wake cycle.

\textbf{Contrast} exhibits pronounced distance-dependence in LS and SWS but not in REM or awake states. This pattern may correspond to the alternating periods of activity and quiescence (resembling UP and DOWN states) that characterize non-REM sleep.

\textbf{Homogeneity} reaches its maximum during wakefulness, indicating a more uniform distribution of E-I co-occurrences, while sleep states show decreased homogeneity (in EIS space). This could reflect different organizational principles of neural activity across vigilance states, with potentially more temporally localized dynamics during sleep (compared to wake) due to the presence of slow oscillation.

\paragraph{Network-level interpretation.}
These multiscale feature patterns suggest potential connections to stochastic network regimes described in theoretical work \citep{Amit1997,Destexhe2006}. While definitive assignments require further investigation, the distinctive features we observe may relate to properties of asynchronous-irregular (AI), synchronous-regular (SR), and asynchronous-regular (AR) dynamics operating in different regions of parameter space. The differential sensitivity of excitatory and inhibitory populations to control parameters—previously observed in Ising model fits \citep{Zanoci2019}—likely contributes to these state-specific feature shifts.

Collectively, these findings suggest that the cortex operates along a critical manifold where transient symmetry breaking occurs within an overall balanced framework. Each vigilance state represents a different region on this manifold, with distinct patterns of E-I coordination that may serve specific computational or homeostatic functions. The dynamic modulation of E-I balance across different sleep states could provide a mechanism for achieving diverse functional outcomes, from information processing during wakefulness to memory consolidation during NREM sleep to novel association formation during REM sleep. The MNCM analysis provides a quantitative approach to characterizing these state-dependent variations, offering a bridge between renormalization-group concepts and physiological network dynamics. Further research integrating modeling with larger datasets will be essential to refine our understanding of how these multiscale features relate to underlying network states and computational processes.

  \begin{widetext}
  \onecolumngrid
\begin{figure}[tp]
\includegraphics[scale=1]
{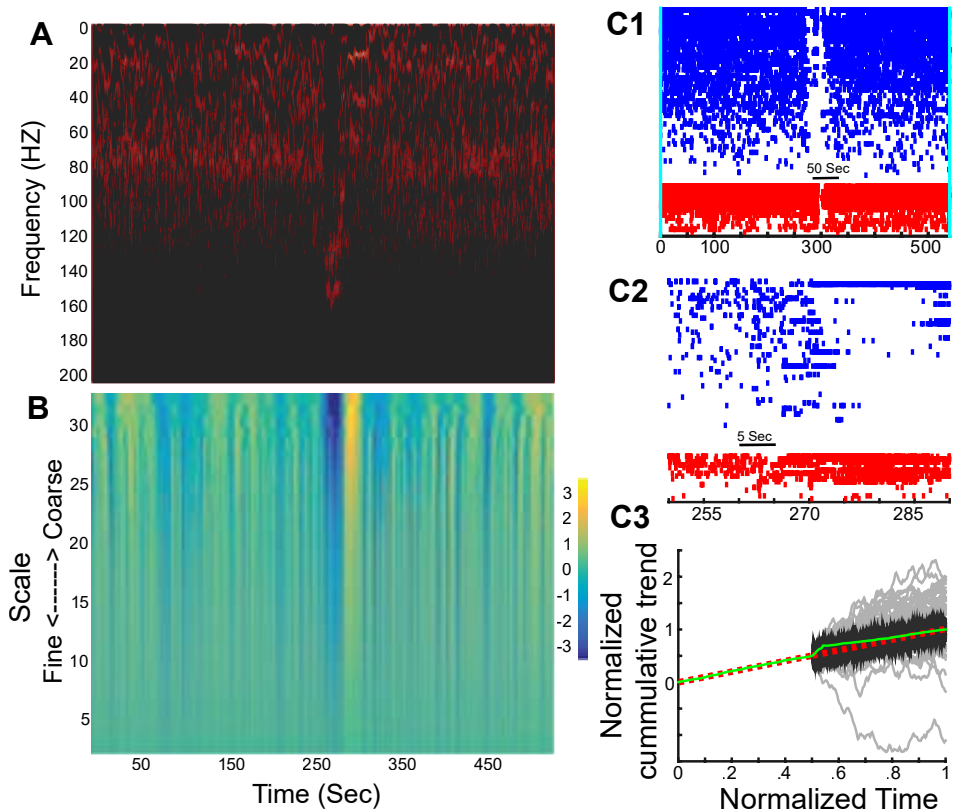}
\caption{
\textbf{A: Local field potential (LFP) multiscale during seizure.} \emph{A1. Multiscale coherence ridges.} Crazy-climber algorithm, a relaxation method based on MCMC, was used to extract multiple ridges from the energetic distribution in LFP wavelet coherence. Sharp transition in activity across scales (frequency ranges) shows the emergence of seizure. \emph{A2. Multiscale heatmap of synaptic current.} Synaptic current (estimated from the normalized ensemble spiking) also shows a multiscale signature of pathological symmetry breaking (during seizure) at the same time that multi-ridges of wavelet coherence show a sharp transition from low to high frequencies.
\textbf{B: Symmetry breaking in pathological state (seizure).} \emph{B1.} Rasterplot of excitatory (blue) and inhibitory (red) unit activity during a 9 minute epoch, \emph{B2.} zoom in to the middle 80 sec. \emph{B3. Reset of balanced activity after seizure.}  Normalized cumulative activity of the signed ensembles (in green) shows a deviation from the estimated balanced trajectory (red) at the onset of seizure. Dark and light gray lines show Monte Carlo simulation of trend stationary and difference stationary processes. The normalized cumulative activity shows a return to the balance trajectory toward the end of the epoch.
}
\label{fig:Lfp} 
\end{figure}
  \twocolumngrid
  \end{widetext}

\subsubsection{Pathological Symmetry Breaking and Post-Ictal Reset}

\paragraph{Breakdown of balance during seizures.}
Our analysis reveals that the finely tuned excitation-inhibition balance breaks down during electrographic seizures \citep{Dehghani2016}. This critical disruption manifests as significant alterations in neuronal firing patterns—inhibitory activity decreases while excitatory firing surges, causing the difference signal $f_{\mathrm{obs}}$ to deviate dramatically from zero. 

Here we show that, simultaneously, the local field potential (LFP) undergoes an abrupt transformation in its oscillatory composition, reflecting macroscopic consequences of this microscopic imbalance. To characterize these dynamic changes, we applied wavelet coherence analysis with ridges extracted using the crazy-climber algorithm \citep{Torrence1998,Carmona1997,Carmona1999}. As shown in Figure \ref{fig:Lfp}A1, seizure onset is marked by a rapid cascade from high- to low-frequency components, aligning with established frequency signatures of ictal events in electrophysiological recordings \citep{Franaszczuk2011,Zaveri2011,Schevon2012,Truccolo2011}. This pattern reveals how seizures fundamentally alter the spectral organization of cortical activity across multiple timescales.

\paragraph{Multiscale signature in reconstructed synaptic current.}
To bridge single-neuron spiking with population-level oscillations, we constructed an estimated synaptic current by convolving spikes with physiologically relevant exponential kernels ($\tau_{E}=3\text{ ms},\;\tau_{I}=10\text{ ms}$) and subtracting inhibitory from excitatory contributions. The resulting time-scale heat map (Fig.\,\ref{fig:Lfp}A2) reveals two pronounced bands of imbalance that align precisely with the LFP coherence ridges: an initial band marking seizure onset, followed by another significant fluctuation that suggests an attempted re-equilibration. These bands delineate the temporal boundaries of the seizure and highlight the multiscale nature of pathological symmetry breaking.


\paragraph{Raster evidence and cumulative deviation.}
The nine-minute raster plot (Fig.\,\ref{fig:Lfp}B1–B2) further illustrates how seizures disrupt the normal coordination between excitatory and inhibitory populations. By calculating the signed cumulative activity 
$\displaystyle S(t)=\sum_{t' \le t}\bigl[\mathrm{Ens}_{E}(t')-\mathrm{Ens}_{I}(t')\bigr]$,
we can quantify the progressive deviation from balance. As shown in Figure \ref{fig:Lfp}B3, the actual trajectory (red line) diverges significantly from the projected balanced path (green line) during the seizure. Monte Carlo simulations of trend-stationary and difference-stationary processes (dark and light gray envelopes, respectively) confirm that this excursion is too large to arise from stochastic fluctuations alone (p < 0.01), indicating a fundamental breakdown in the underlying regulatory mechanisms.

\paragraph{Post-ictal balance reset.}
Perhaps the most remarkable finding is that $S(t)$ gradually returns to the pre-seizure trajectory within approximately 200 seconds after the seizure terminates. This reversion suggests an active network mechanism that can restore the global excitation-inhibition set-point following extreme disturbances. Such adaptive reset could involve multiple homeostatic processes, including synaptic depression of excitatory connections, upregulation of inhibitory efficacy, or activation of neuromodulatory feedback pathways that respond to elevated activity levels.

While these patterns were observed consistently across six focal seizures from two subjects, a larger cohort will be necessary to establish the generality of this ``balance homeostat'' and to dissect its underlying mechanisms across different seizure types and etiologies.

Together, the LFP coherence ridges, synaptic-current heat map, and cumulative-activity trajectory provide a comprehensive view of pathological symmetry breaking and subsequent recovery. They illustrate how the cortex can be displaced from its scale-free balanced state, transit through a pathological regime characterized by profound excitation-inhibition imbalance, and ultimately return toward the renormalization-group-like fixed point that governs normal cortical dynamics—a remarkable demonstration of the brain's capacity for self-regulation even after extreme departures from homeostasis.

\section{Discussion: Synthesis}\label{sec:discussion}

\subsection{Why a Renormalization Mind-Set, and Not Just Mean-Field?}

Macroscopic descriptions often expose simpler, more universal regularities than the intricate, microscopic interactions from which they emerge \cite{Bulla2008}.  In neuroscience, two strategies embody this philosophy.  
\emph{Neural field} models keep space explicit, treating activity as a continuous function over cortical sheets and capturing wave-like propagation and pattern formation \citep{WilsonCowan1972, WilsonCowan1973, Byrne2020}.  
\emph{Neural mass} models collapse space altogether, tracking a small set of lumped variables—typically the average membrane potential or firing rate of a homogeneous pool—to reproduce bulk rhythms and evoked responses \citep{Coombes2006, Cook2022}.  
Both frameworks rest on a core mean-field assumption: heterogeneous interactions are replaced by an ``average'' field that each unit feels (impacted by).  While this step yields analytically tractable equations and has clarified many dynamical motifs \citep{Pintosis2014, Robinson2021}, it also imposes three restrictive simplifications:

1. \textbf{Gaussian fluctuations.} Higher-order, non-Gaussian statistics—ubiquitous in spiking data—are projected out.  
2. \textbf{Linear response bias.} Perturbations are usually treated in a linear (or weakly nonlinear) regime, masking the strong, state-dependent nonlinearities that dominate near critical points.  
3. \textbf{Single-scale view.} Coupling across temporal or spatial scales is ignored, even though cascades ranging from millisecond spikes to slow population waves shape cortical computation \citep{Tiberi2022}.

Renormalization group (RG) techniques, originally developed in statistical physics to handle multiscale phenomena near critical points, address precisely the kinds of limitations that mean-field approaches face. The core idea is to \emph{systematically integrate out} fast or small-scale degrees of freedom—those ``high-momentum'' modes—and then rescale the remaining variables so that the coarse-grained system can be compared with its former self \citep{Wilson1971_b, Fisher1998, Meurice2011, Wilson1975}. Fixed points of this iterative procedure signal scale invariance; the flow of system parameters toward or away from such points quantifies how correlations and nonlinear couplings evolve across scales.  In neural data, verifying that the probability distribution of coarse-grained activity is (approximately) invariant provides an operational test for criticality \citep{Meshulam_2019}.  Phenomenological RG applied to single-unit recordings in mouse hippocampus, for instance, revealed power-law scaling consistent with a supercritical, finite-size regime \citep{Nicoletti2020}.  

By adopting an RG-inspired temporal coarse-graining of collective excitatory and inhibitory spike trains, we aim to retain those multiscale, non-Gaussian interactions for heterogenous populations that conventional mean-field (neural mass or field) models average away.  This perspective supplies a principled bridge between microscopic spiking variability and emergent macrodynamics, offering new traction on where—and how—the cortex sits relative to criticality.

\subsection{Scale-Invariant Ensemble E/I Balance: Key Findings and Their Implications}
\label{subsec:scale_inv}

\paragraph{Structured correlation beyond statistical averaging.}
Our temporal coarse-graining analysis revealed remarkable scale invariance in the ensemble difference signal $f_{\mathrm{obs}}(t;\Delta t)=\mathrm{Ens}_{E}-\mathrm{Ens}_{I}$ across four orders of magnitude ($10^{-3}$ s to $10^1$ s). As shown in the results, this balance exhibits a sublinear variance scaling relationship, with variability decreasing more slowly than the linear relationship ($\propto 1/\Delta t$) that would be predicted by the central limit theorem for independent processes. This departure from central-limit behavior indicates structured correlations between excitation and inhibition that persist across temporal scales. The consistency of these patterns across wake, REM, and all stages of NREM sleep (Fig.\,\ref{fig:Data}C; Fig.\,\ref{fig:NoiseCollapse}A1; Table \ref{tab:Table1}) suggests that this coordinated balance represents a fundamental organizing principle of cortical dynamics rather than a state-specific phenomenon.

\paragraph{Universal collapse curves and scale invariance.}
The scale invariance of E-I balance is further demonstrated by our collapse-curve analysis, which shows that rescaled distributions of $f_{\mathrm{obs}}$ converge onto a single ``master'' curve with remarkable precision across both temporal scales and behavioral states (Fig.\,\ref{fig:NoiseCollapse}B). This universal collapse pattern represents a defining empirical signature of a renormalization group fixed point, where the system's statistical properties remain invariant under successive coarse-graining transformations. The consistency of this pattern across four decades of temporal resolution indicates that the underlying mechanisms maintaining E-I balance operate through similar principles regardless of the timescale of observation.

\paragraph{Randomization control tests.}
Our randomization tests provide crucial evidence that the observed scale invariance emerges from precise temporal coordination rather than from statistical artifacts. By disrupting timing relationships while preserving firing rate distributions, both ISI permutation and circular shift methods transform the structured, elliptical E-I relationship into isotropic clouds and destroy the collapse-curve universality. These results  establish that balanced E-I dynamics depend fundamentally on structured correlations—both within and between neurons—across multiple timescales, rather than simply matching mean firing rates between populations.

\paragraph{Partition-curve geometry and microstate accessibility.}
The partition curve $P(p;\Delta t)$—which measures the fraction of spikes contained in the top $p\%$ of time bins—revealed another dimension of scale invariance. These curves remain strongly convex at every scale, indicating that even after extensive coarse-graining, neural firing remains highly inhomogeneous, with a small minority of time bins carrying the majority of spikes. From a statistical mechanics perspective, while coarse-graining necessarily integrates out degrees of freedom and reduces entropy, it never fully homogenizes the distribution: burst-like dynamics survive even macroscopic averaging. The surrogate partition curves, in contrast, migrate rapidly toward the diagonal with increasing bin size, confirming that the empirical inhomogeneity reflects genuine, temporally coordinated ensemble activity rather than random fluctuations.

\subsection{Symmetry Breaking Across Vigilance States}

\paragraph{From ensemble balance to functional flexibility.}
Distinguishing between \textit{input} E/I balance \citep{Renart2010,vanVreeswijk1996} and the \textit{ensemble} balance we explore is essential. While input balance ensures stable single-cell firing rates, ensemble balance reveals macroscopic organizational principles across temporal scales. In our framework, `symmetry breaking' refers to transient disruptions in this ensemble equilibrium—unlike static physical systems where symmetry breaking leads to new stable states, cortical networks dynamically restore balance after transient imbalances, enabling computational flexibility.

\paragraph{State-dependent dynamics on a critical manifold.}
Our multidimensional feature analysis through MNCM reveals that different vigilance states represent distinct patterns of symmetry breaking within an overall balanced framework. These state-specific signatures suggest the cortex operates along a critical manifold where each state occupies a different region with characteristic properties. The wakeful state exhibits uniform E-I co-occurrences that may optimize flexible information processing, while sleep states show more heterogeneous patterns potentially corresponding to the temporally localized dynamics of slow oscillations. REM sleep displays a unique intermediate pattern that may support internally-driven processing during dreams.

These dynamic modulations of ensemble E-I balance likely reflect fundamental changes in cortical operating modes that serve specific computational functions—from information processing during wakefulness to memory consolidation during NREM sleep to novel association formation during REM. Previous studies have demonstrated how fine-scale balance regulates spike timing through coordinated excitatory-inhibitory conductances \citep{Higley2006,Rudolph2007}, with inhibitory activity crucial for maintaining cortical excitability \citep{Trevelyan2005,Pouille2013}. Our approach extends these insights to the population level, bridging theoretical concepts of symmetry breaking with physiological network dynamics to explain how the brain achieves different functional states while maintaining overall stability.

\subsection{Pathological Symmetry Breaking in Seizures}

\paragraph{Seizures as extreme deviations on the critical manifold.}
While normal brain states represent different regions on a critical manifold with controlled symmetry breaking, seizures constitute severe excursions from this manifold. Our analyses reveal that these pathological events involve a multiscale breakdown of balance - from altered single-unit firing patterns to profound shifts in population-level oscillations across multiple frequency bands. This multiscale nature of seizures is evident in both the temporal domain (spanning milliseconds to minutes) and across measurement modalities (from spike trains to local field potentials), representing a system-wide collapse of the carefully maintained E/I equilibrium.

The most striking finding is the active re-establishment of multiscale balance after such profound disruption. The statistically significant return of cumulative activity to the pre-seizure trajectory (compared to stochastic models) suggests the existence of a ``balance homeostat'' operating over timescales of tens of seconds and less than a couple of minutes. This homeostatic mechanism likely engages multiple processes spanning synaptic, cellular, and network levels—potentially including activity-dependent depression of excitatory connections, compensatory inhibitory upregulation, or neuromodulatory feedback.

From a theoretical perspective, seizures represent a transient escape from the critical manifold that governs normal cortical function. The consistent return to this manifold after extreme perturbation reinforces the view that balanced excitation-inhibition is not merely an epiphenomenon but a dynamically maintained target state essential for information processing. This conceptualization of seizures as temporary deviations from a critical manifold, rather than transitions to distinct attractors, offers a novel framework for understanding both epileptogenesis and post-ictal recovery.

\subsection{Relation to Other Frameworks of Criticality}

\paragraph{Beyond traditional critical point theories.}
Our findings challenge conventional frameworks of neural criticality by emphasizing the crucial distinction between excitatory and inhibitory populations—a separation typically overlooked in traditional approaches. The invariance of probability distributions under repeated coarse-graining provides a rigorous test for criticality beyond thermodynamic analogies \citep{Meshulam_2019}. While traditional self-organized criticality (SOC) models predict a single control-parameter-free attractor and power-law avalanche distributions, our data reveal a more complex picture: scale-invariant E/I balance persists across different brain states and withstands undesirable perturbations, suggesting a critical regime that extends beyond a single point in parameter space.

\paragraph{Limitations of existing models and evidence for a critical manifold.}
This perspective builds directly on our previous work demonstrating that neuronal avalanches do not conform to SOC predictions when rigorously analyzed \citep{Dehghani2012}. In that study, we showed that avalanche size distributions from units and local field potentials in cat, monkey, and human cortex during wake and sleep states did not exhibit true power-law scaling—instead following bi-exponential distributions when subjected to proper statistical testing. This finding directly challenged the prevailing SOC framework that fails to account for the role of inhibition and the complexity of spatiotemporal patterns in neuronal networks.

Similarly, our work on pairwise maximum entropy (Ising) models revealed that ignoring inhibitory neurons dramatically overestimates synchrony among excitatory neurons, and that excitatory and inhibitory populations show distinct properties in their collective behavior \citep{Zanoci2019}. The present study extends these insights by explicitly tracking the differential contributions of excitation and inhibition across scales, revealing dynamics that cannot be captured by treating neural populations as homogeneous.

Multiple lines of evidence support our proposed critical manifold:

The MNCM analysis reveals distinctive state-specific signatures across vigilance states, showing how the brain can occupy different regions of a critical manifold while maintaining overall balance. Each state exhibits unique patterns of Energy, Entropy, Contrast, and Homogeneity that likely serve specific computational functions, yet collapse curves from these same states converge onto remarkably similar regimes. This simultaneous presence of state-specific features and universal scaling behavior resembles critical phenomena in physical systems \citep{Papanikolaou2011}.

Our partition curve analysis demonstrates persistent inhomogeneity in firing distributions even after extensive coarse-graining—a finding incompatible with simple statistical averaging but consistent with an intrinsically structured heterogeneity maintained across scales. This preserved burst-like structure represents another dimension of invariance that complements the collapse curve results.

Our randomization tests provide crucial validation of these findings. When timing relationships are disrupted while preserving rate distributions, both the partition curve inhomogeneity and collapse curve universality disappear, confirming that these properties depend on precise temporal coordination rather than statistical artifacts. Interestingly, at only the coarsest scales, surrogate collapse curves begin to approach data curves, suggesting that while overall balance is maintained, fine-scale information structure diminishes at extended timescales—consistent with observations in avalanche dynamics where information content decreases at larger scales \citep{Friedman2012}.

\paragraph{Theoretical implications.}
Rather than a single critical point, our data indicate an extended critical manifold stabilized by recurrent inhibition, with multiple ``marginal'' directions along which perturbations neither grow nor decay exponentially. This criticality emerges from competing interactions between distinct excitatory and inhibitory populations—a heterogeneity that is fundamental to understanding cortical dynamics yet frequently overlooked in homogeneous models of neural criticality.

This framework unifies our observations across normal and pathological conditions: different vigilance states represent distinct regions along the critical manifold, while seizures constitute temporary excursions away from this manifold followed by homeostatic return. The persistent inhomogeneity revealed by partition curves, the state-dependent signatures in MNCM features, and the universal collapse curves all point to a system that maintains balance through active coordination of distinct neural populations rather than through a homogeneous self-organizing mechanism.

By explicitly modeling excitatory and inhibitory contributions separately and examining their relationship across scales, we reveal principles of cortical organization that reconcile stability with flexibility—maintaining overall balance while allowing for state-specific modulations and transient deviations that support information processing.

\subsection{Implications and Future Directions}

Our framework of scale-invariant balance with state-dependent symmetry breaking offers three key insights:
\begin{enumerate}
\item \textbf{Dynamic adaptability through balanced criticality:} The brain maintains E/I balance across scales while permitting transient, state-specific deviations that enable both stability and flexibility. This balance appears to be actively maintained rather than passively emerging, suggesting cortical networks have evolved specialized homeostatic mechanisms operating across multiple timescales.

\item \textbf{Multiscale constraints on neural architecture:} The persistence of E/I balance from milliseconds to seconds places strong design constraints on mechanistic models of cortical function. Models that treat excitation and inhibition as homogeneous or neglect cross-scale interactions will miss the organizing principles revealed by our analysis.

\item \textbf{Understanding pathological recovery:} The post-ictal recovery trajectory demonstrates the brain's remarkable capacity to restore balance after extreme perturbations, providing a window into homeostatic mechanisms that normally operate within narrower bounds during healthy function.
\end{enumerate}

While current multielectrode arrays sample only a small fraction of neurons within a cortical column, advancements in recording technology will enable more comprehensive sampling of neural populations. Future research integrating cellular neurophysiology with network-level models will be crucial for identifying the specific mechanisms that implement the balance homeostat and potentially revealing new therapeutic approaches focused on stability enhancement rather than mere excitability reduction.

\subsection{Conclusion}

As Philip Anderson noted in his seminal paper ``More is Different'', large systems might not exhibit the symmetry of their governing laws, and temporal ordering alongside broken symmetry are pivotal in biological systems \cite{Anderson1972}. Our findings suggest that cortical networks operate at what we term the \emph{edge of symmetry}—a critical manifold where excitation and inhibition remain balanced across timescales, yet small, structured symmetry breakings enable flexible computation.

Unlike systems like crystals or magnets where symmetry breaking leads to minimum energy states, the cortex dynamically restores balance after transient imbalances, sacrificing energy minimization for computational flexibility. This parallels Prigogine's concept of symmetry breaking in dissipative structures \citep{Prigogine_1967,Prigogine_1968}, where biological systems harvest energy to return to unstable symmetric states. We speculate that the significant energy consumption of the cortex may be largely devoted to re-establishing this unstable symmetry that is constantly breaking during information processing. Like other dissipative structures, the cortex exhibits multistability, sustained oscillations, and propagating waves—all signatures of a system that maintains dynamic organization far from equilibrium \citep{Goldbeter_2018,Prigogine_1969}.

By integrating renormalization-group concepts with neural dynamics, we provide a framework for understanding how the cortex maintains functional stability while enabling adaptive information processing across states from normal waking and sleep to pathological seizures. The multiscale nature of this balance supports a computational regime that is simultaneously stable yet flexible—allowing cortical circuits to process inputs over several orders of magnitude without driving the network away from equilibrium \citep{Xing1996}.

Let us emphasize that the complexity of neural activity patterns, emerging from fundamental principles like scale-invariant balance, highlights the contrast between the simplicity of the underlying laws and their diverse outcomes. By quantifying the catalog of invariances across states and (spatial and temporal) scales, we can search for governing principles and diverse neural activity patterns that emerge from these simple laws.

 \begin{acknowledgments}
The author thanks prior collaborators, Alain Destexhe and Max Tegmark for their valuable inputs. This work was not supported by any funding.
 \end{acknowledgments}

%
%
%
%


\section*{References}
\bibliography{symmetry}

\end{document}